\documentclass[aps,nofootinbib,twocolumn,prd,eqsecnum,showpacs,showkeys,preprintnumbers]{revtex4-1}

\usepackage[caption=false]{subfig}
\usepackage{graphicx}
\usepackage{amsmath}
\usepackage{amsfonts}
\usepackage{amssymb}
\usepackage{color}
\usepackage{bm}
\usepackage{mathrsfs}
\usepackage{epstopdf}
\usepackage{url}
\usepackage{footnote}
\usepackage{textcomp}

\makeatletter
\newcommand*{\rom}[1]{\expandafter\@slowromancap\romannumeral #1@}
\makeatother

\begin{document}

\title{Scalar perturbations from brane-world inflation with curvature effects}
\author{Mariam Bouhmadi-L\'{o}pez $^{1,2}$}
\email{mariam.bouhmadi@iem.cfmac.csic.es}
\author{Pisin Chen $^{3,4,5,6}$}
\email{chen@slac.stanford.edu}
\author{Yen-Wei Liu $^{3,5}$}
\email{f97222009@ntu.edu.tw}
\date{\today}

\affiliation{
${}^1$Instituto de Estructura de la Materia, IEM-CSIC, Serrano 121, 28006 Madrid, Spain\\
${}^2$Centro Multidisciplinar de Astrof\'{\i}sica - CENTRA, Departamento de F\'{\i}sica, Instituto Superior T\'ecnico, Av. Rovisco Pais 1,1049-001 Lisboa, Portugal\\
${}^3$Department of Physics, National Taiwan University, Taipei, Taiwan 10617\\
${}^4$Graduate Institute of Astrophysics, National Taiwan University, Taipei, Taiwan 10617\\
${}^5$Leung Center for Cosmology and Particle Astrophysics, National Taiwan University, Taipei, Taiwan 10617\\
${}^6$Kavli Institute for Particle Astrophysics and Cosmology, SLAC National Accelerator Laboratory, Stanford University, Stanford, CA 94305, U.S.A.
}

\begin{abstract}
We consider a generalization of the Randall-Sundrum single brane-world  scenario (RS2). More precisely, the generalization is described through curvature corrections, corresponding to a Gauss-Bonnet term in the bulk and a Hilbert-Einstein term, as well as the strength of the induced gravity term, on the brane. We are mainly interested in analyzing the early inflationary era of the brane, which we model within the extreme slow-roll limit, i.e., under a de Sitter  like brane inflation,  where the inflaton field is confined on the brane. We compute the scalar perturbations in this model and compare our results with those previously obtained for the RS2 scenario with and without an induced gravity term on the brane or a Gauss-Bonnet term in the bulk. The amplitude of the scalar perturbations is decreased as compared with a pure RS2 model. In addition, the effect from the Gauss-Bonnet correction in an induced gravity brane-world model is to decrease the amplitude of the scalar perturbations, and a similar result is obtained for the induced gravity effect in a Gauss-Bonnet brane-world. In general, in the high energy limit the amplitude is highly suppressed by the Gauss-Bonnet effect. Finally, we constrain the model using the latest WMAP7 data.
\end{abstract}

\maketitle

\section{introduction}
Several approaches in particle physics and cosmology imply the possibility that our observable universe is a hypersurface embedded in a higher-dimensional space-time, which is motivated by superstring/M theory. In this scenario several brane-world models have been proposed (cf. for example Ref.~\cite{Maartens:2010ar}). One of the most popular and interesting brane-world scenario is provided by the Randall-Sundrum single brane model (RS2) \cite{hep-th/9906064}, where our universe corresponds to a four-dimensional (4d) single brane embedded in a five-dimensional (5d) anti de-Sitter (AdS$_5$) bulk. In this brane-world model, matter fields are localized on the brane, and only the gravitons can propagate into the bulk. Although the extra dimension is infinite in RS2 model, the zero-mode of the 5d graviton is localized on the brane at the low energy limit due to the warped geometry of the bulk metric, which corresponds to the 4d gravitational waves, and therefore this property allows the RS2 brane-world to recover standard general relativity (GR) at the low energy limit. By contrast, at the high energy limit the RS2 model gives an unconventional modification to standard GR.

The 5d bulk action in the RS2 model is given by a Hilbert-Einstein action, which can be generalized by adding some correction terms. With regard to the 4d cosmology, there are two important modifications to the RS2 model; the first one is the Gauss-Bonnet (GB) correction term included in the bulk action, which leads to the most general second-order field equation in a 5d bulk \cite{Lovelock:1971yv}. Moreover, this unique combination of the GB term in the bulk action also corresponds to the leading corrections in string theory, and it is a ghost-free combination \cite{Zwiebach:1985uq,Zumino:1985dp}. Besides, it also plays an important role in Chern-Simons gravity \cite{Witten:1988hc,Chamseddine:1989nu,Zanelli:2005sa}, which is a gauge theory of gravity. Furthermore, the zero-mode of the 5d gravitons in the GB brane-world is also localized on the brane at low energy as in the RS2 model \cite{Deruelle:2003tz}. The second modification is the induced-gravity (IG) correction added in the brane action. This effect is generated due to the quantum loops of matter fields on the brane that couple to the bulk gravitons~\cite{Zakharov:1970cc,Collins:2000yb,Dvali:2000hr,Dvali:2000xg,Shtanov:2000vr, Sahni:2002dx}.

Here we investigate the RS2-type model modified by both the GB and IG effects \cite{Kofinas:2003rz,BouhmadiLopez:2008nf}. In addition, we are mainly interested in the early inflationary era of the brane, where the spatial curvature and the dark radiation term are rapidly diluted. Furthermore, in order to compare with the power spectrum of the scalar perturbations in the RS2 model, we consider the normal branch of the model which recovers the standard GR at the low energy limit without an effective cosmological constant on the brane. In addition, this branch reduces to the RS2 model in the absence of GB and IG corrections to the bulk and brane gravitational action (Please see the first reference in \cite{BouhmadiLopez:2008nf}).

When applying the RS2 brane-world to the early universe cosmology, if we consider single-field inflation localized on the brane in the extreme slow-roll limit, there is no scalar zero-mode contribution from the bulk; moreover, the massive KK scalar mode projected from the bulk metric can be neglected since they are too heavy to be excited during inflation \cite{Langlois:2000ns,hep-ph/9912464}. Therefore, the formula for the power spectrum of the scalar perturbations in this approximation is the same as that based on GR (see Eqs.(\ref{As2}) and (\ref{standard})). We assume that this remains valid on our case.

The scalar perturbations on brane-world inflation in the extreme slow-roll limit have been investigated in Ref.~\cite{hep-ph/9912464} (RS2 model), Refs.~\cite{BouhmadiLopez:2004ax} and \cite{Papantonopoulos:2004bm} (RS2 model with IG effect) and Ref.~\cite{Dufaux:2004qs} (pure GB brane-world). When the GB and IG corrections are both included, we show that the effect from the GB correction in an IG brane-world model is to decrease the amplitude of the scalar perturbations, and a similar result is obtained for the IG effect in a GB brane-world. The same effects have been obtained for the pure RS2 model with either a GB bulk term or an IG brane term \cite{BouhmadiLopez:2004ax,Dufaux:2004qs}. Before concluding, we would like to point out that the scalar perturbations for a GB brane-world model with a Dvali-Gabadadze-Porrati (DGP) action on the brane, i.e., a particular case of induced gravity and which has been proposed to describe the late-time evolution of the universe, has been previously analyzed in Ref.~\cite{Nozari:2011zz}. Here we handle the most general case by including all kind of curvature corrections on the brane and in the bulk. In addition, our generalization is fully in agreement with RS2 model, i.e., the brane-world model we consider contain the RS2 kind of fine-tuning; that is the brane is flat in the absence of matter due to a balance between the brane tension, the bulk cosmological constant and the curvature effects. The later mentioned model is suitable to describe the early time evolution of the universe unlike the first mentioned model.

The outline of the paper is as follows. In section \rom{2}, we consider the RS2 kind of action modified by a GB correction in the 5d bulk action as well as an IG correction on the brane action, and we review how to obtain the modified Friedmann equation on the brane. In section \rom{3}, we solve the modified Friedmann equation analytically and pick up the normal branch of the general solutions, i.e., the branch that reduces to the RS2 model when GB and IG corrections vanish. In section \rom{4}, we calculate the amplitude of the scalar perturbations of this kind of brane inflation in the extreme slow-roll limit and then we constrain the model  using WMAP7 data \cite{Komatsu:2010fb}. Finally, in section \rom{5} we present our conclusions and discussions.

\section{the model}
We consider a 5d brane-world model where the brane split the bulk into two symmetric pieces. The bulk action contains a GB term in addition to the usual Hilbert-Einstein term, while the brane action is described by an IG term, a brane tension and a Lagrangian for matter. Then the action of the system is given by \cite{footnote1}:
\begin{align}
S=&\frac1{2\kappa_5^2}\int_Md^5x\sqrt{-g}\left[R-2\Lambda_5+\alpha\left(R^2-4R_{\mu\nu}R^{\mu\nu}+\right.\right.\notag \\
  &\left.\left.R_{\mu\nu\rho\sigma}R^{\mu\nu\rho\sigma}\right)\right]+\int_{\partial M}d^4x\sqrt{-h}\left[\frac{\gamma}{2\kappa^2_4}\hat{R}-\lambda+\mathscr{L}_m\right]\label{action},
\end{align}
where $g_{\mu\nu}$ and $h_{\mu\nu}$ are the bulk metric and  the brane metric, respectively, $\kappa_5^2$  is the bulk gravitational constant, $\Lambda_5(\leq0)$ is the bulk cosmological constant, $\alpha(\geq0)$ the GB parameter which has the dimension of length square, $\gamma$ is a dimensionless parameter indicating the strength of the IG term, and $\lambda$ is the brane tension.

On the one hand, the GB term in Eq.(\ref{action}) is a ghost-free combination of higher-order curvature tensors, which also corresponds to the leading quantum corrections in string theory; moreover, it plays an important role in Chern-Simons gravity~\cite{Witten:1988hc,Chamseddine:1989nu,Zanelli:2005sa}. On the other hand, the IG correction is generated due to the quantum loops of matter fields on the brane which couple to the bulk gravitons~\cite{Zakharov:1970cc,Collins:2000yb,Dvali:2000hr,Dvali:2000xg,Shtanov:2000vr,Sahni:2002dx}.

The effective Einstein equation on the brane can be deduced by two means: (i) solving the bulk and junction conditions simultaneously, either by using some specific coordinates~\cite{Binetruy:1999ut,Binetruy:1999hy,Kofinas:2003rz} or through a covariant approach~\cite{Shiromizu:1999wj,Maeda:2003ar,Maeda:2003vq} (ii) choosing an appropriate bulk solution and imposing the junction condition at the brane. For clarity and simplicity, we will briefly review in the next paragraphs the second approach~\cite{Davis:2002gn}.

We consider the static uncharged black hole solution in 5d GB gravity~\cite{Boulware:1985wk,Cai:2001dz,Banados:1993ur,Crisostomo:2000bb}:
\begin{equation}
ds^2=-f(r)dT^2+f^{-1}(r)dr^2+r^2\Omega_{ij}dx^idx^j,
\end{equation}
where
\begin{equation}
f=k+\frac{r^2}{4\alpha}\left(1\pm\sqrt{1+\frac43\alpha\Lambda_5+8\alpha\frac{\tilde{\mu}}{r^4}}\right).\label{bulksol}
\end{equation}
The parameter $\tilde\mu$ is related to the black hole mass. There are two branches for the black hole solutions. However, we disregard the ``$+$'' branch as it is unstable. The reason behind this instability is that the graviton degree of freedom is a ghost. In addition,  the mass of the black hole is negative in this branch~\cite{Boulware:1985wk,Cai:2001dz}.

A few words in relation to the junction conditions on this kind of model are worthy: the junction conditions are essentially the same as in GB brane-world gravity without IG \cite{Davis:2002gn,Maeda:2003vq,Charmousis:2002rc,Maeda:2007cb}. The presence of an IG term results in a shift of the total energy momentum tensor of the brane, which appears in the junction condition.
For the bulk black hole solution (\ref{bulksol}), we impose the junction condition at the brane and we obtain the effective 4d Friedmann equation~\cite{{Kofinas:2003rz},{footnote2}}:
\begin{align}
&\left[1+\frac83\alpha\left(H^2+\frac{k}{a^2}+\frac{\Phi}2\right)\right]^2\left(H^2+\frac{k}{a^2}-\Phi\right)\notag\\
&=\frac{\gamma
^2}4\left(\frac{\kappa_5}{\kappa_4}\right)^4\left[H^2+\frac{k}{a^2}-\frac{\kappa^2_4}{3\gamma}\left(\rho+\lambda\right
)\right]^2\label{Friedmann},
\end{align}
where
\begin{equation}
\Phi+2\alpha\Phi^2=\frac{\Lambda_5}6+\frac C{a^4},
\label{eqphi}
\end{equation}
and $k=0,\pm 1$. The constant $C$ is related to the mass of the bulk black hole, and it measures the strength of the dark radiation on the brane that is inversely proportional to $a^4$. The condition (\ref{eqphi}) results in two possible values for $\Phi$
\begin{equation}
\Phi_\pm=\frac1{4\alpha}\left[-1\pm\sqrt{1+8\alpha\left(\frac{\Lambda_5}6+\frac C{a^4}\right)}\;\right].
\label{phipm}
\end{equation}
From now on, we will restrict our analysis to the solution with $\Phi_+$, %which for simplicity we will denote simply $\Phi$.
because it corresponds to the stable bulk solution with ``$-$'' sign in Eq.(\ref{bulksol}), moreover, we recover a Hilbert-Einstein action in the bulk if $\alpha\rightarrow 0$ and therefore the model we are considering reduces to the RS2 scenario in the absence of any curvature corrections of the GB and IG kind. This is very important because one of our main aims in this paper is to see how the amplitude of the scalar perturbations on RS2 model are modified by including GB and IG terms. For simplicity, from now on we will drop the subscript ``$+$'' on $\Phi_+$.

We are mainly interested in the early inflationary era of the brane, where the spatial curvature and the dark radiation are quickly washed  out. Therefore, from now on we will consider a spatially flat brane ($k=0$) within an AdS$_5$ bulk ($C=0$).

By imposing the RS2 kind of fine-tuning, i.e., the effective cosmological constant on the brane vanishes; in other words in the absence of matter on the brane the Hubble rate vanishes, we obtain
\begin{equation}
\frac{\kappa^4_5}{36}\lambda^2=-\Phi\left(1+\frac34\alpha\Phi\right)^2\label{fine},
\end{equation}
which implies that $\Phi$ is negative. For later convenience, we introduce a new positive variable $\mu=\sqrt{|\Phi|}$, then Eq.(\ref{fine}) can be rewritten as
\begin{equation}
\kappa^2_5\lambda=2\mu\left(3-4\alpha\mu^2\right)\label{fine-tuning}.
\end{equation}
Therefore, the parameter $\mu^2$ is bounded as \mbox{$0\leq\mu^2<1/4\alpha\,$} (see also Ref.~\cite{footnote2-a}), as can be then easily deduced by using Eq.(\ref{phipm}) (for $k=0$ and $C=0$) and assuming the natural requirement of a positive brane tension.

\section{the branches of the Friedmann equation}\label{branches}

In order to proceed further, we need first to solve the cubic Friedmann equation (\ref{Friedmann}). This equation was previously analyzed in Refs.~\cite{Kofinas:2003rz} and \cite{Brown:2006mh}. We will solve this equation analytically extending the methodology used in Ref.~\cite{BouhmadiLopez:2008nf}. It is important to solve this equation analytically because it will allow us to pick up the right branch, i.e., the branch that reduce to RS2 model in the absence of curvature corrections of the sort IG and GB,  for analyzing the scalar perturbations on the brane in the extreme slow-roll limit.

This analysis can be simplified by taking the square root of both sides of the generalized Friedmann equation:
\begin{align}
&\left[1+\frac83\alpha\left(H^2-\frac{\mu^2}2\right)\right]\sqrt{H^2+\mu^2}\notag\\
&=\pm \,r_c\left[\frac{\kappa^2_4}{3}\left(\rho+\lambda\right)-\gamma H^2\right]\label{sqrt},
\end{align}
where $r_c\equiv\kappa^2_5/2\kappa^2_4$. Now we choose the ``+'' sign because it contains the RS2 solution for $\alpha\rightarrow0$ and $\gamma\rightarrow0$, as we will show later (see also Ref.~\cite{footnote3}). In addition, to solve Eq.(\ref{sqrt}) analytically, it is more convenient to introduce the dimensionless parameters~\cite{BouhmadiLopez:2008nf}:
\begin{align}
&\bar{X}=\frac{8}3\frac{\alpha}{\gamma r_c}\sqrt{H^2+\mu^2}\label{barX},\\
&b=\frac{8}3\frac{\alpha}{\gamma^2 r_c^2}\left(1-4\alpha\mu^2\right),\\
&\bar{\rho}=\frac{64}9\frac{\alpha^2}{\gamma^2 r_c^2}\mu^2+\frac{32}{27}\frac{\alpha^2\kappa^2_5}{\gamma^3r_c^3}\left(\rho+\lambda\right)\label{rhobar}.
\end{align}
Then the Friedmann equation (\ref{sqrt}) can be rewritten in a simpler form:
\begin{equation}
\bar{X}^3+\bar{X}^2+b\bar{X}-\bar{\rho}=0\label{cubic}.
\end{equation}
Notice that $\bar X$ is positive and therefore only positive solutions of Eq.(\ref{cubic}) are physically meaningful.

The number of the real solutions of the cubic equation (\ref{cubic}) depend on the sign of the discriminant $N$ \cite{Abramowitz}:
\begin{equation}
N=Q^3+R^2,
\end{equation}
where
\begin{equation}
\left\{
\begin{array}{ll}
\displaystyle Q=\frac13\left(b-\frac13\right),\\
\displaystyle R=\frac16b+\frac12\bar{\rho}-\frac1{27}\,\,.
\end{array}
\right.
\end{equation}
It is more convenient to factorise the discriminant $N$ as
\begin{equation}
N=Q^3+R^2=\frac14(\bar{\rho}-\bar{\rho}_1)(\bar{\rho}-\bar{\rho}_2),
\label{discriminant}
\end{equation}
where
\begin{align}
\bar{\rho}_1&=-\frac13\left\{b-\frac29\left[1+\sqrt{(1-3b)^3}\right]\right\}, \label{rho1}\\ \bar{\rho}_2&=-\frac13\left\{b-\frac29\left[1-\sqrt{(1-3b)^3}\right]\right\}.
\label{rho2}
\end{align}
If $N$ is positive, there is only one real solution. If $N$ is negative, all solutions are real, and if $N$ vanishes, all solutions are real and at least two of them are equal.

%%%%%%%%%%%%%%%%%%%%%%%%%%%%%%%%
\begin{figure}[!h]
\centering
\includegraphics[width=6cm]{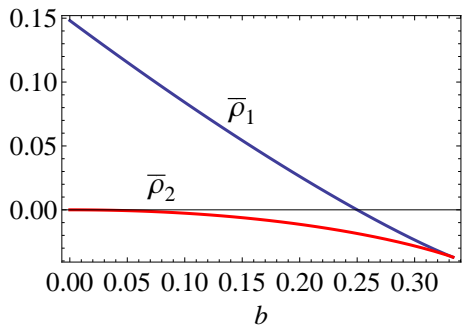}
\caption{This plot shows the behavior of the dimensionless functions $\bar\rho_1$ and $\bar\rho_2$ versus $b$ (see Eqs~(\ref{rho1})-(\ref{rho2})).}
\label{rho}
\end{figure}
%%%%%%%%%%%%%%%%%%%%%%%%%%%%%%%%

The functions $\bar{\rho}_1$ and $\bar{\rho}_2$ are defined in terms of the parameter  $b$ (see Eqs.(\ref{rho1})-(\ref{rho2}) and Fig.~\ref{rho}). The signs and values acquired by   $\bar{\rho}_1$ and $\bar{\rho}_2$ are very important as for a given dimensionless parameter $\bar\rho$ and for a given value of $b$, the sign of the discriminant is fully determined by these two functions (cf. Eq.(\ref{discriminant})). We can then split our analysis into four cases depending on the sign of $N$ and the value of $b$:
\begin{itemize}

\item If $0\leq b<1/4$, $\bar{\rho}_1$ is positive; while $\bar{\rho}_2$ is negative and vanishes when $b=0$. Therefore we have: (i) $N$ is positive when $\bar\rho_1<\bar\rho$, where there is a unique real solution of Eq.(\ref{cubic}). (ii) $N$ is negative when $\bar\rho_2<\bar\rho<\bar\rho_1$, allowing for three real solutions of Eq.(\ref{cubic}). (iii) $N$ vanishes if $\bar\rho=\bar\rho_1$ or $\bar\rho=\bar\rho_2=0$. In this special case, there are three real solutions and at least two of them are equal.

\item If $1/4\leq b<1/3$, $\bar{\rho}_1$ and $\bar{\rho}_2$ are negative ($\bar\rho_1$ vanishes when $b=1/4$). Here: (i) $N$ is strictly positive as long as $0<\bar\rho$, consequently, allowing for a unique real solution of  Eq.(\ref{cubic}). (ii) $N$ can vanish if and only if (iff) $b=1/4$ and $\bar\rho=0$. For this fine tuned case, there are three real solutions where at least two are equals.

\item If $b=1/3$, $\bar{\rho}_1$ and $\bar{\rho}_2$ are equal and negative. Then $N$ is positive and again there is a unique real solution of Eq.(\ref{cubic}).

\item If $1/3 <b$, $\bar{\rho}_1$ and $\bar{\rho}_2$ are complex conjugate and therefore $N$ is positive for any energy density and there is a unique real solution of the modified Friedmann equation (\ref{cubic}).

\end{itemize}

Note that we are solving the modified Friedmann equation (\ref{cubic}) without imposing the generalized fine-tuning condition \textit{\`{a} la} Randall-Sundrum as imposed by Eq.(\ref{fine-tuning}). The constraint (\ref{fine-tuning}) will be imposed when calculating the spectrum of the scalar perturbations on the next section.

We next solve the modified Friedmann equation for the different cases stated above bearing in mind that $\bar X$ must be positive (cf. Eq.(\ref{barX})).

\subsection{Case 1: $0\leq b<\frac14$}
\begin{figure}[h]
\centering
  \includegraphics[width=6.5cm]{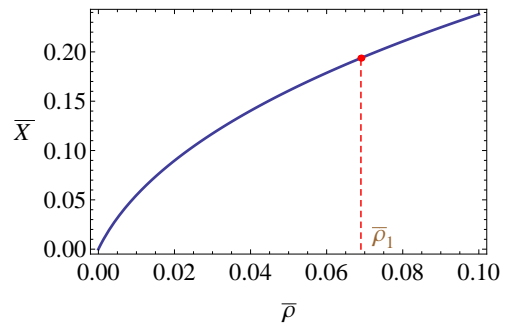}
	\caption{This figure shows the positive solution of the modified Friedmann equation (\ref{cubic}) for $0\leq b<1/4$. More precisely, we plot $\bar X$, essentially the Hubble rate (see Eq.(\ref{barX})), versus the dimensionless energy density $\bar\rho$. For this plot, we set $b=1/8$, as an example. Notice that the solutions (\ref{high}) and (\ref{intermediate}) are continuous at $\bar\rho=\bar\rho_1$ through the solution (\ref{limiting}).}
	\label{case1}
\end{figure}

In this region $\bar{\rho}_1$ is positive and $\bar{\rho}_2$ is negative ($\bar{\rho}_2$ can vanish when $b=0$). Furthermore, we can divide the set of solutions into three different cases depending on the sign of the discriminant $N$. After doing a careful analysis of the cubic Friedmann equation (\ref{cubic}), the only positive root of that equation when $0\leq b<1/4$ reads (see Fig.~\ref{case1}):
\begin{align}
&\bar{X}=\frac13\left[2\sqrt{1-3b}\cosh{\left(\frac{\eta}{3}\right)}-1\right]\label{high};&&
0<\bar{\rho}_1<\bar{\rho}, \\
&\bar{X}=\frac13\left[2\sqrt{1-3b}\cos{\left(\frac{\theta}{3}\right)}-1\right]\label{intermediate};&&
0<\bar{\rho}<\bar{\rho}_1, \\
&\bar{X}=\frac13\left[2\sqrt{1-3b}-1\right]\label{limiting};&&
\bar{\rho}=\bar{\rho}_1,
\end{align}
where the parameters $\eta$ and $\theta$ are defined as follows
\begin{align}
&\cosh{\eta}=\frac{R}{\sqrt{-Q^3}}\,\,,  &&\sinh{\eta}=\sqrt{\frac{Q^3+R^2}{-Q^3}}\,\,,\label{eta1} \\
&\cos{\theta}=\frac{R}{\sqrt{-Q^3}}\,\,, &&\sin{\theta}=\sqrt{\frac{Q^3+R^2}{Q^3}}\,\,,\label{theta}
\end{align}
which gives the constraint $0<\eta$. The dimensionless parameter $\bar{\rho}$ is always positive (see Eq.(\ref{rhobar})), and therefore there is an upper bound for the angle $\theta$:
\begin{equation}
\theta_{max}=\arccos{\left(\frac{-2+9b}{2\sqrt{(3b-1)^3}}\right)}.
\end{equation}
In fact, in general we have a constraint on the angle for these set of solutions: $0<\theta\leq\theta_{max}$. If the dimensionless energy density $\bar{\rho}=\bar{\rho}_2=0$ which is the case iff $b=0$, we obtain the boundary values for $\bar X$:
\begin{equation}
\bar{X}=\left.\frac13\left[\sqrt{1-3b}-1\right]\right|_{b=0}=0.
\end{equation}

In the very early universe, i.e., at high energy, the parameter $\eta$ goes to infinity when the energy density gets very large. On this regime, the Friedmann equation can be approximated by:
\begin{equation}
H\sim\frac12\left(\frac{\kappa^2_5}{2\alpha}\right)^{\frac13}\rho^{\frac13}\label{highenergy}.
\end{equation}
It is worthy to stress that this modified Friedmann equation is different from the one obtained in: (i) standard general relativity ($H\sim\sqrt{\rho}$), (ii) RS2 model ($H\sim\rho$) and (iii) induced gravity ($H\sim\sqrt{\rho}$) \cite{Maeda:2003ar,hep-th/0408061} (on the last two cases we are referring to the high energy regime). But coincide with that of a pure GB brane-world model (see for example \cite{Dufaux:2004qs,astro-ph/0303168}).

We highlight that in general there are three possible branches for $\bar X$, but two of them are negative, and thus we do not include them in Fig.~\ref{case1}. Before concluding this subsection we would like to point out that the solutions (\ref{high}) and (\ref{intermediate}) are continuous at $\bar\rho=\bar\rho_1$ through the solution (\ref{limiting}).

\subsection{Case 2: $\frac14\leq b<\frac13$}
In this branch the function $\bar{\rho}_1$ and $\bar{\rho}_2$ are negative and $\bar{\rho}_1$ vanishes when $b=1/4$ . Moreover, we can split this branch into two different cases depending on the sign of the discriminant $N$. If the dimensionless energy density $0<\bar\rho$, the only positive solution of Eq.(\ref{cubic}) is
\begin{equation}
\bar{X}=\frac13\left[2\sqrt{1-3b}\cosh{\left(\frac{\eta}{3}\right)}-1\right]\label{B},
\end{equation}
where $\eta$ is defined in Eq.~(\ref{eta1}), consequently, we set $\eta>0$. The high energy limit of the solution (\ref{B}) also has the approximate form given in Eq.(\ref{highenergy}).

In addition, for the limiting case of $\bar{\rho}=\bar{\rho}_1=0$ where the parameter $b=1/4$, we obtain the boundary solution:
\begin{equation}
\bar{X}=\left.\frac13\left[2\sqrt{1-3b}-1\right]\right|_{b=\frac14}=0.
\end{equation}

\subsection{Case 3: $b=\frac13$}
For this special choice of the parameter $b=1/3$, $\bar{\rho}_1$ and $\bar{\rho}_2$ acquires the same negative value. The discriminant is strictly positive for any given dimensionless energy density $\bar\rho$, and thus there is a unique real solution which reads:
\begin{equation}
\bar{X}=\frac13\left[\left(1+27\bar{\rho}\right)^{\frac13}-1\right].\label{C}
\end{equation}
The high energy approximation in the early universe is the same shown on the previous cases (see Eq.(\ref{highenergy})).

\subsection{Case 4: $\frac13<b$}
When the parameter $b$ is larger than $1/3$, the functions $\bar{\rho}_1$ and $\bar{\rho}_2$ become a pair of complex conjugates, and therefore the discriminant $N$ is strictly positive for any dimensionless parameter $\bar\rho$, then there is a unique real solution in this branch which can be written as:
\begin{equation}
\bar{X}=\frac13\left[2\sqrt{3b-1}\sinh{\left(\frac{\xi}3\right)}-1\right],
\label{D}
\end{equation}
where $\eta$ is defined as
\begin{equation}
\sinh{\xi}=\frac{R}{\sqrt{Q^3}}\,\,,\,\,\, \cosh{\xi}=\sqrt{\frac{Q^3+R^2}{Q^3}}\,\,,
\end{equation}
with the constraint $\xi>0$. The same high energy approximation applies to this case in the early universe (see Eq.(\ref{highenergy})):

\subsection{Recovering RS2 solution}

We now show that the solution (\ref{intermediate}) has a well-defined limit when $\alpha\rightarrow0$, which reduces to the normal branch in the RS2 model with an IG correction. Furthermore, this approximation reduces to the RS2 model when we take as well the limit $\gamma\rightarrow0$.

The parameters $\bar X$, $b$ and $\bar\rho$ depend on the GB parameter $\alpha$ with different power. Therefore, it is convenient to introduce the following definition when doing the expansion around $\alpha=0$ \cite{BouhmadiLopez:2008nf}:
\begin{align}
&\bar X=f_1\alpha;         &&f_1=\frac83\frac1{\gamma r_c}\sqrt{H^2+\mu^2}\,\,,\\
&b=f_2\alpha+f_3\alpha^2;  &&f_2=\frac83\frac1{\gamma^2 r^2_c},\,\, f_3=-\frac{32}3\frac{\mu^2}{\gamma^2 r^2_c},\\
&\bar\rho=f_4\alpha^2;     &&f_4=\frac{64}9\frac{\mu^2}{\gamma^2 r^2_c}+\frac{32}{27}\frac{\kappa^2_5}{\gamma^3 r^3_c} (\rho+\lambda),
\end{align}
where $f_1$, $f_2$, $f_3$ and $f_4$ do not depend on the value of the GB parameter $\alpha$.
%%%%%%%%%%%%%%%%%%%%%%%%%%%%%%%%%%%%%%%%%%%%%%%
\begin{figure*}[!ht]
  \centering
  \subfloat[]{\label{G2}\includegraphics[width=0.35\textwidth]{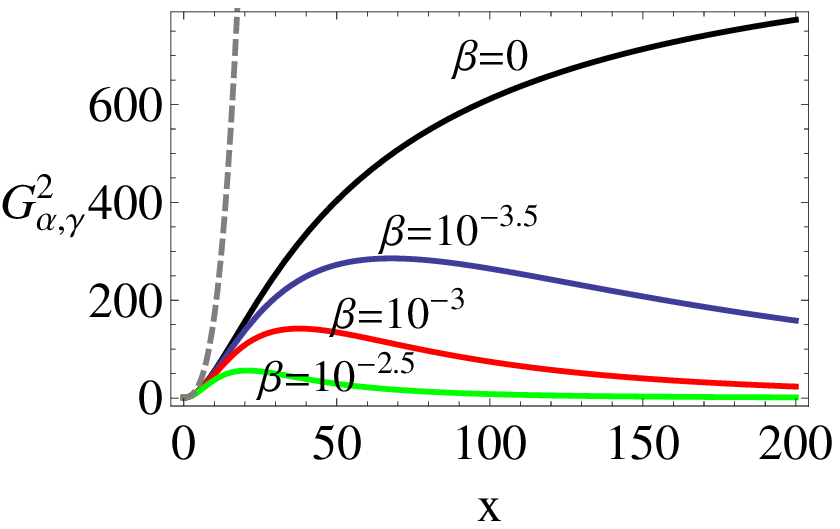}}\qquad~
  \subfloat[]{\label{G1}\includegraphics[width=0.35\textwidth]{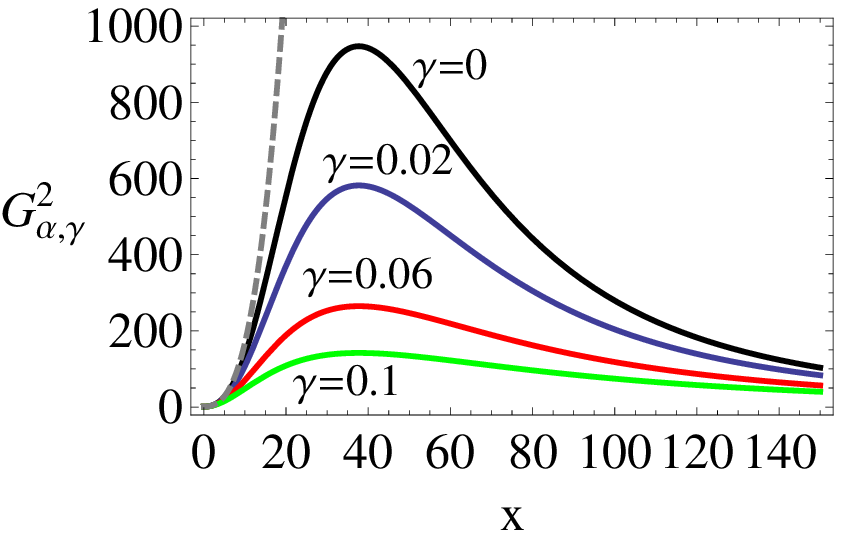}}
  \caption{The amplitude of the scalar perturbations normalized to the standard 4d results in the extreme slow-roll limit (see Eq.(\ref{As}) and Eq.(\ref{G})) against the scale of inflation on the brane normalized to the square root of the absolute value of the effective cosmological constant in the bulk. In figure (a), as an example we have fixed the IG parameter, more precisely, $\gamma=0.1$, and changed the GB parameter $\beta$ as shown on the plot
. In figure (b), as an example we have fixed the GB parameter, more precisely, $\beta=4\alpha\mu^2=10^{-3}$, and changed the IG parameter $\gamma$ as shown on the plot. We can see that the effect of a GB term in the bulk action of an IG brane-world model is to decrease the amplitude of the scalar perturbations. Similarly, we can see that the effect of IG on a GB brane-world model is to decrease the amplitude of the scalar perturbations. The same effect applies for the RS2 model under the presence of a GB bulk term and/or an IG brane term \cite{BouhmadiLopez:2004ax,Dufaux:2004qs}.}
  \label{Gcorrection}
\end{figure*}
%%%%%%%%%%%%%%%%%%%%%%%%%%%%%%%%%%%%%%%%%%%%%%%%
We notice that If the GB effect approaches to zero (when $\alpha\rightarrow0$), the parameter $b\rightarrow0$ and the dimensionless energy density $\bar\rho\rightarrow0$. Therefore, here we only analyze the solution (\ref{intermediate}) under this approximation.

The angle $\theta$ in the solution (\ref{intermediate}) is defined in Eq.(\ref{theta}) and it goes to $\pi$ when $\alpha$ goes to zero. Then we can do the following expansion:
\begin{equation}
\sin(\pi-\theta)=\left(\sqrt{\frac{27}4f_2^2+27f_4}\,\,\right)\alpha+O(\alpha^2),
\end{equation}
Thus, we obtain the relation
\begin{equation}
\theta\approx\pi-\left(\sqrt{\frac{27}4f_2^2+27f_4}\,\,\right)\alpha+O(\alpha^2),
\end{equation}
therefore, we get the expansion
\begin{equation}
\cos\left(\frac{\theta}3\right)\approx\frac12+\frac12\left(\sqrt{\frac94f_2^2+9f_4}\,\,\right)\alpha+O(\alpha^2).
\end{equation}
We then substitute the above result into Eq.(\ref{intermediate}) with the definition of $f_1$, $f_2$ and $f_4$, and we obtain the expansion equation
\begin{equation}
f_1\alpha=\frac12\left(-f_2+\sqrt{f^2_2+4f_4}\right)\alpha+O(\alpha^2).\label{lowalpha}
\end{equation}
We then consider the lowest order expansion of the GB parameter $\alpha$ in Eq.(\ref{lowalpha}), and notice that the parameter $\mu^2\simeq-\Lambda_5/6$ when $\alpha\rightarrow0$. Finally, we get a well-defined result in this approximation:
\begin{align}
H^2&=\frac16\frac{\kappa^2_5}{\gamma r_c}(\rho+\lambda)+\frac1{2\gamma^2 r^2_c} \notag\\
   &-\frac1{2\gamma^2 r^2_c}\sqrt{1+\frac23\gamma r_c\left[\kappa^2_5(\rho+\lambda)-\gamma r_c\Lambda_5\right]}\,\,\,,
   \label{RSIG}
\end{align}
which is the normal branch of the RS2 model with the IG correction term. Furthermore, if we take the limit $\gamma\rightarrow0$ (which is the limit when the effect from the IG correction vanishes) in the Eq.(\ref{RSIG}), we obtain
\begin{equation}
H^2=\frac{\kappa^4_5\lambda}{18}\rho\left(1+\frac{\rho}{2\lambda}\right)+\frac{\Lambda_5}{6}+\frac{\kappa_5^4\lambda^2}
{36},
\label{RS}
\end{equation}
which is the RS2 model without the GB and IG corrections. In summary, the solution (\ref{intermediate}) has a well-defined approximation given by Eq.(\ref{RSIG}) when $\alpha\rightarrow0$, which corresponds to the normal branch in RS2 model with IG correction. Besides, it recovers the RS2 model Eq.(\ref{RS}) when we take the limit $\alpha\rightarrow0$ and $\gamma\rightarrow0$.

\section{scalar perturbations on the brane}
%%%%%%%%%%%%%%%%%%%%%%%%%%%%%%%%%%%%%%%%%%%%%%%
\begin{figure*}[!ht]
  \centering
  \subfloat[]{\label{ratiogammafixed}\includegraphics[width=0.4\textwidth]{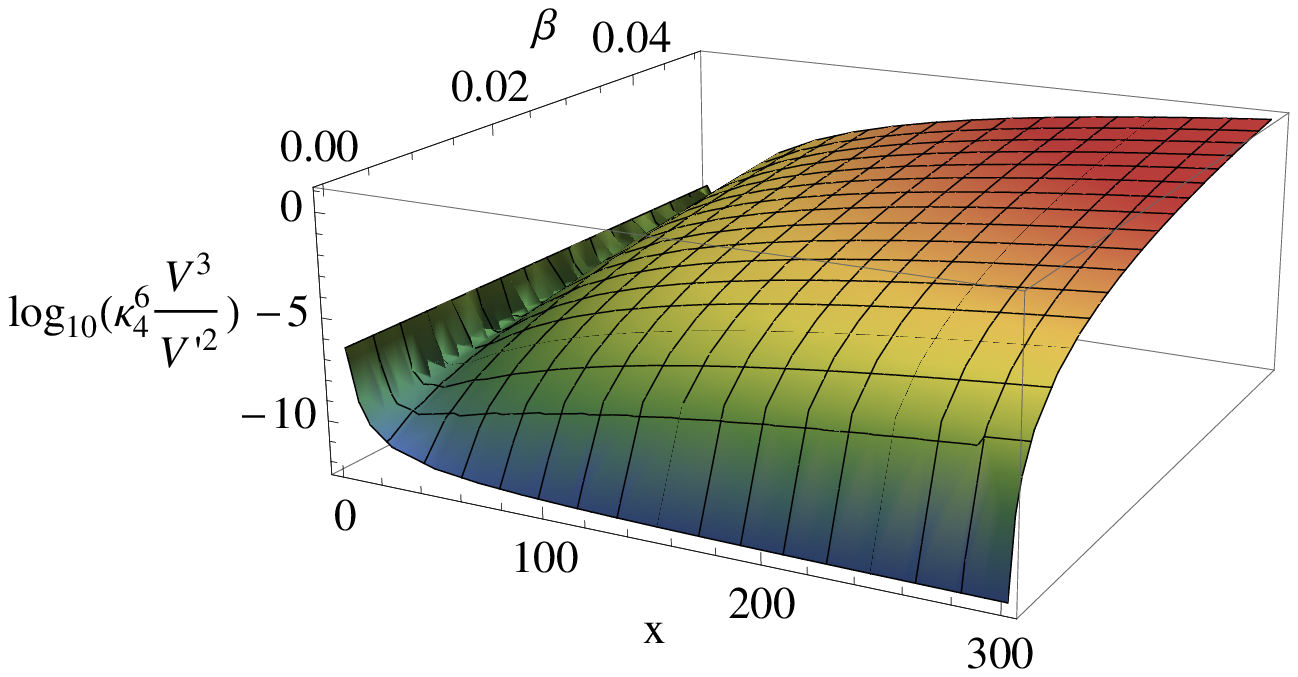}}  \qquad
  \subfloat[]{\label{ratiobetafixed}\includegraphics[width=0.4\textwidth]{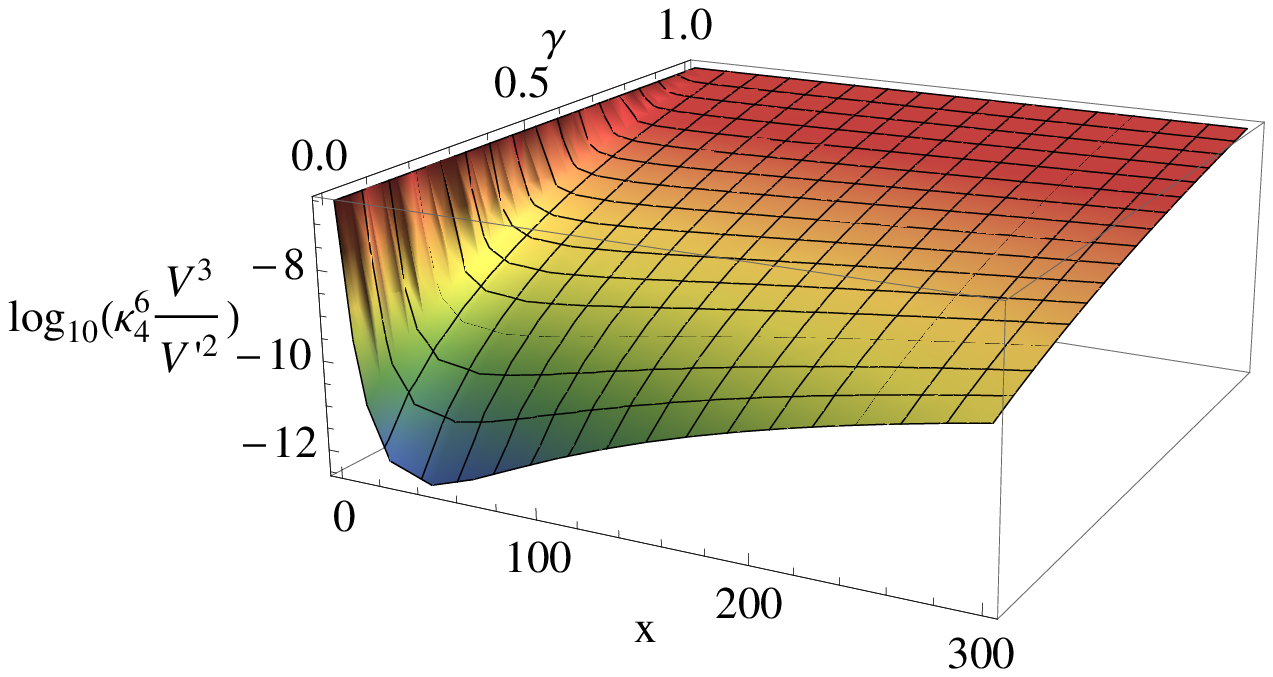}}
	\caption{\label{ratioconstraint} We constrain the inflaton potential by means of the amplitude of the scalar perturbations as measured by WMAP7, i.e., $P_s = 2.45 \times 10^{-9}$ at the pivot scale $k_0 = 0.002$ Mpc$^{-1}$. More precisely, in plot (a) we show $\log_{10}(\kappa_4^6V^3/V'^2)$ versus the scale of inflation, $x=H/\mu$, and the dimensionless GB parameter, $\beta$, for a fixed value of the dimensionless IG parameter, $\gamma$, such that $\gamma=0.1$; in plot (b) we show $\log_{10}(\kappa_4^6V^3/V'^2)$ versus the scale of inflation, $x=H/\mu$, and the dimensionless IG parameter, $\gamma$, for a fixed value of the dimensionless GB parameter, $\beta$, such that $\beta=10^{-3}$.}
\end{figure*}
%%%%%%%%%%%%%%%%%%%%%%%%%%%%%%%%%%%%%%%%%%%%%%%
In this section we consider single field inflation on the brane and study the lowest order of the scalar perturbations in the extreme slow-roll limit. Within this approximation, we assume there is no additional scalar zero-mode contribution from the bulk while the massive scalar mode from the bulk are too heavy to be excited during the inflationary era which is true in RS2 brane-world models \cite{Langlois:2000ns}. Therefore, the massive KK scalar mode can be neglected in the extreme slow-roll inflation. With these assumptions one can follow the standard procedure to calculate the power spectrum for the scalar perturbations \cite{Lidsey:1995np}, and we will follow the procedures for brane inflation used in \cite{hep-ph/9912464,Dufaux:2004qs,BouhmadiLopez:2004ax}.

Before calculating the power spectrum of the scalar perturbations, we need first to find out the 4d effective gravitational constant on the brane, which is an essential ingredient when calculating the scalar perturbations. Therefore, we focus on the generalized RS2 solutions modified by GB and IG effects and impose the generalized fine-tuning condition given in Eq.(\ref{fine-tuning}).

We note that the solutions of the generalized Friedmann equation we discussed in the previous section (see Eqs.(\ref{high})-(\ref{intermediate}), Eq.(\ref{B}), Eq.(\ref{C}) and Eq.(\ref{D})) are the real solution (\ref{1stsol}) of the general cubic equation given in the appendix \ref{appendix1}. It is more convenient to analyze the solution (\ref{1stsol}) in the low energy limit (when $\rho\rightarrow0$) to find the 4d effective gravitational constant.

In the low energy region when $\rho\rightarrow0$, we first impose the RS2 kind of fine-tuning condition expressed by Eq.(\ref{fine-tuning}) on the dimensionless energy density (\ref{rhobar}), and then do the series expansion of the solution Eq.(\ref{1stsol}) around $\rho=0$, then we obtain
\begin{equation}
\frac{8}3\frac{\alpha}{\gamma r_c}\sqrt{H^2+\mu^2}\approx\frac{8}3\frac{\alpha\mu}{\gamma r_c}+\frac{8\mu\kappa^2_5}{9(1+2\gamma r_c\mu+4\alpha\mu^2)}\rho+O(\rho^2).
\end{equation}
By squaring both sides of the above expression, we obtain
\begin{equation}
H^2\approx\frac{\mu\kappa^2_5}{3(1+2\gamma r_c\mu+4\alpha\mu^2)}\rho+O(\rho^2).\label{lowenergy}
\end{equation}
We notice that this approximation recovers standard GR without an ``extra'' cosmological constant on the brane as should be for RS2 inspired models.
The coefficient in front of the energy density defines the 4d gravitational constant where we remaind that the crossover scale, $r_c$, is defined as $r_c\equiv\kappa_5^2/2\kappa_4^2$. Therefore, we find the relation between the effective gravitational constant on the brane  $\kappa_4^2$ and the bulk gravitational constant $\kappa_5^2$:
\begin{equation}
\kappa_4^2=\left(\frac{1-\gamma}{1+4\alpha\mu^2}\right)\mu\kappa^2_5\label{graviconst}.
\end{equation}

The relation between $\kappa^2_4$ and $\kappa^2_5$ implies: (i) $\gamma$ is bounded; $0\leq\gamma<1$, (ii) $\mu$ must be strictly positive, i.e., $\mu>0$.
We can take the limit $\alpha\rightarrow0$ and $\gamma\rightarrow0$ on  Eq.(\ref{graviconst}), and we find that the approximated result is consistent with Ref.~\cite{BouhmadiLopez:2004ax} ($\alpha\rightarrow0$, pure IG case), Ref.~\cite{Dufaux:2004qs} ($\gamma\rightarrow0$ pure GB case) and Ref.~\cite{hep-th/9906064} ($\alpha\rightarrow0$, $\gamma\rightarrow0$, i.e., RS2 case). On the other hand, this result is different from the one in Refs.~\cite{Kofinas:2003rz} and \cite{gr-qc/0508116}. That is because the authors in Refs.~\cite{Kofinas:2003rz} and \cite{gr-qc/0508116} considered a specific GB brane-world model ($\gamma=1$) without imposing the RS2 kind of fine-tuning condition given by Eq.(\ref{fine-tuning}).
%%%%%%%%%%%%%%%%%%%%%%%%%%%%%%%%%
\begin{figure*}[!ht]
  \centering
  \subfloat[]{\label{C1gamma}\includegraphics[width=0.256\textwidth]{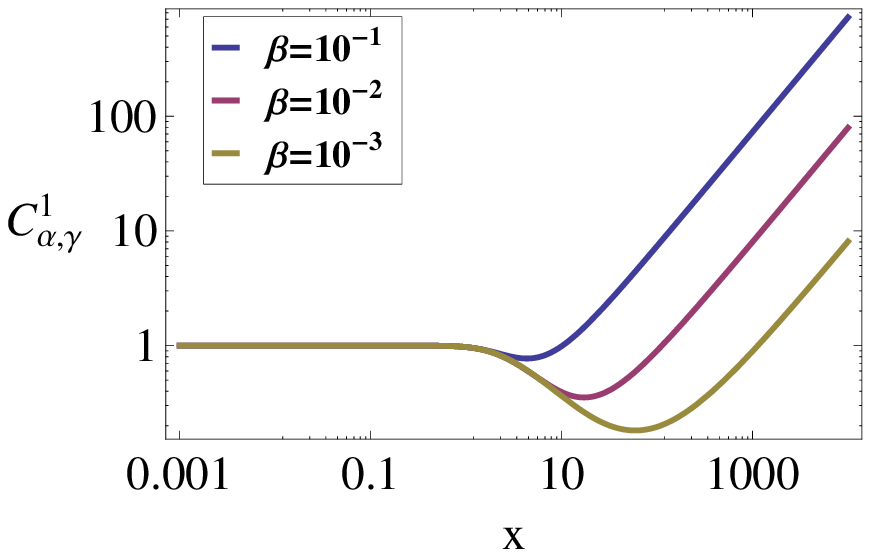}}
  \subfloat[]{\label{C2gamma}\includegraphics[width=0.256\textwidth]{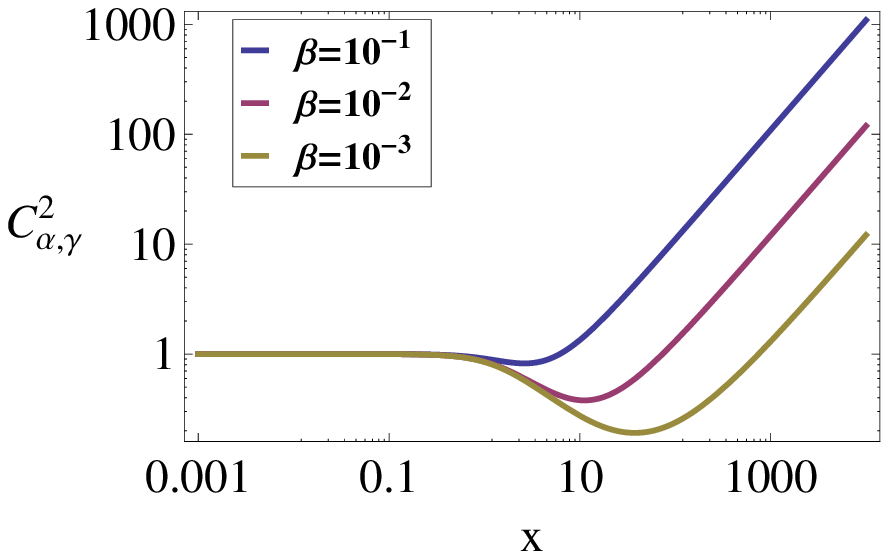}}
  \subfloat[]{\label{C1beta}\includegraphics[width=0.256\textwidth]{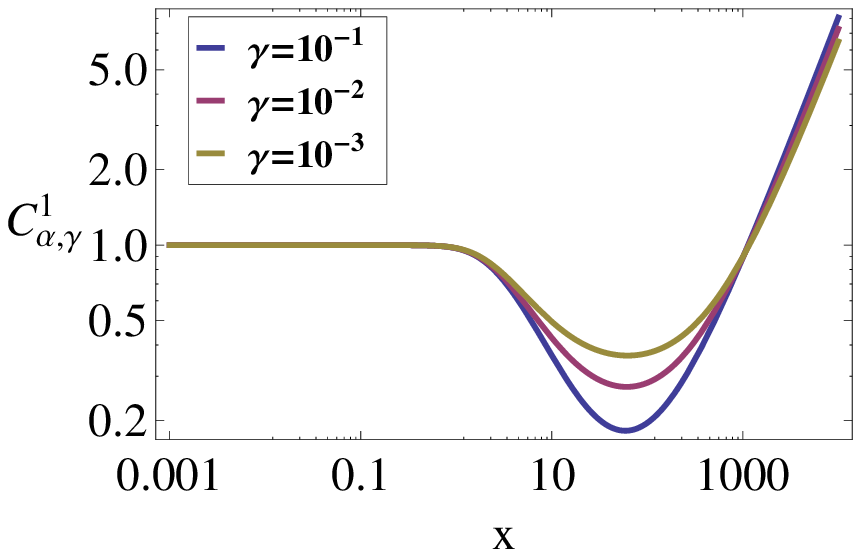}}
  \subfloat[]{\label{C2beta}\includegraphics[width=0.256\textwidth]{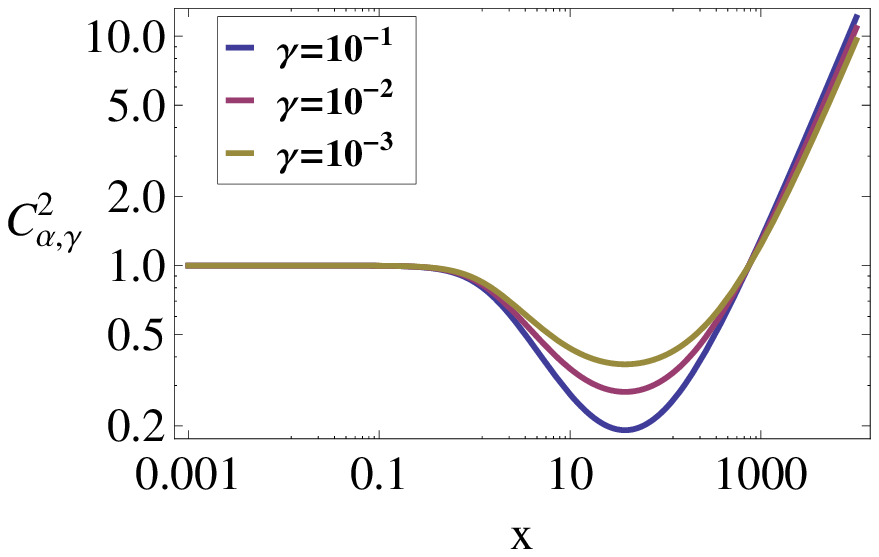}}
	\caption{The corrections to the standard 4d slow-roll parameters $\epsilon_{4D}$ and $\eta_{4D}$ (see Eq.(\ref{epsilon}) and Eq.(\ref{eta})) versus the dimensionless energy scale $x=H/\mu$. In figures (a) and (b), we have fixed the IG parameter, $\gamma=0.1$, and changed the GB parameter $\beta$ as shown on the plots; in figures (c) and (d), we have fixed the GB parameter, $\beta=10^{-3}$, and changed the IG parameter $\gamma$ as shown on the plots.}
	\label{slowroll}
\end{figure*}
%%%%%%%%%%%%%%%%%%%%%%%%%%%%%%%%%

Now, we proceed to calculate the power spectrum of the scalar perturbations. The normalized amplitude of the scalar  perturbations for a given mode that reenter the horizon after inflation is given by \cite{Lidsey:1995np}
\begin{equation}
A_S^2=\frac4{25}\langle\zeta^2\rangle=\frac{H^4}{25\pi^2\dot{\phi}^2}.
\label{As2}
\end{equation}

In addition, in the extreme slow-roll inflation, $\dot{\phi}\simeq-V'(\phi)/3H$; we can therefore substitute this approximation into the amplitude of the scalar perturbations Eq.~(\ref{As2}), resulting in
\begin{equation}
A_S^2=\frac9{25\pi^2}\frac{H^6}{V'^2},
\label{standard}
\end{equation}
which is independent of the gravitational field equation \cite{astro-ph/0003278}. In order to compare conveniently with the standard 4d general relativity results, we rewrite Eq.(\ref{standard}) as:
\begin{equation}
A_S^2=\frac{\kappa^6_4}{75\pi^2}\left(\frac{V^3}{V'^2}\right)G^2_{\alpha,\gamma}=
[A^2_S]_{4\textrm{D}}G^2_{\alpha,\gamma}\label{As},
\end{equation}
where $[A^2_S]_{4\textrm{D}}$ is the standard 4d result and the correction term $G^2_{\alpha,\gamma}$ is defined as
\begin{equation}
G^2_{\alpha,\gamma}=\frac{27H^6}{V^3\kappa^6_4}.
\label{correction}
\end{equation}

In the extreme slow-roll limit, the energy density of the inflaton $\rho\approx V$. We then calculate the potential $V$ in this approximation using Eq.(\ref{sqrt}) (with ``$+$'' sign) as well as the RS2 kind of fine-tuning condition Eq.(\ref{fine-tuning}) and the 4d effective gravitational constant defined in Eq.(\ref{graviconst}). Consequently, we obtain \cite{clarification}
\begin{align}
V&=\frac1{\kappa^2_5}\left\{\left[6+16\alpha\left(H^2-\frac{\mu^2}2\right)\right]\sqrt{H^2+\mu^2}\right.\notag\\
 &+\left.3\gamma H^2\frac{1+4\alpha\mu^2}{\mu(1-\gamma)}-2\mu(3-4\alpha\mu^2)\right\}.
\label{V}
\end{align}
Then we substitute it into the expression (\ref{correction}) and we use as well Eq.(\ref{graviconst}). Finally, we obtain the correction term to the standard 4d result (see also Fig.~{\ref{Gcorrection}}):
\begin{widetext}
\begin{equation}
G^2_{\alpha,\gamma}(x)=\left[\frac{3(1+\beta)x^2}{2(1-\gamma)(3-\beta+2\beta x^2)\sqrt{1+x^2}+3\gamma x^2(1+\beta)+2(1-\gamma)(\beta-3)}\right]^3\label{G},
\end{equation}
\end{widetext}
where $x=H/\mu$ and $\beta=4\alpha\mu^2$. This result is consistent with Ref.~\cite{BouhmadiLopez:2004ax} when $\beta\rightarrow0$ and consistent with Ref.~\cite{Dufaux:2004qs} when $\gamma\rightarrow0$.

In Fig.~{\ref{Gcorrection}}-(a) and Fig.~{\ref{Gcorrection}}-(b) the dashed-grey lines correspond to the amplitude of the RS2 model without GB and IG corrections, which is monotonically increasing with respect to the dimensionless energy scale $H/\mu$. If the GB and IG corrections are both included, we see that the effect from the GB correction in an IG brane-world model is to decrease the amplitude of the scalar perturbations, and a similar result is obtained for the IG effect in a GB brane-world (cf. Fig.~\ref{Gcorrection}). The same effects have been obtained for RS2 model with either GB \cite{Dufaux:2004qs} or IG \cite{BouhmadiLopez:2004ax} effect.

In the very low energy limit, i.e., the Hubble parameter $H\ll\mu$ or $x\rightarrow0$, the correction to the standard 4d result corresponds to $G^2_{\alpha,\gamma}\sim1$. Therefore, the amplitude of the scalar perturbations recovers the 4d standard result. During the intermediate energy scale, the amplitude of the scalar perturbations is enhanced with respect to the energy scale; while in the very high energy regime, i.e., the Hubble parameter $\mu\ll H$ or $x\rightarrow\infty$, we obtain the following approximation:
\begin{equation}
G^2_{\alpha,\gamma}\sim\frac{27}{64}\left[\frac{1+\beta}{\beta(1-\gamma)}\right]^3\frac1{x^3},
\end{equation}
which means that in the high energy limit the perturbation will be highly suppressed by the GB effect.

Finally, we impose observational constraints by using the latest WMAP7 data \cite{Komatsu:2010fb}; for the power spectrum of the scalar perturbations: $P_s$ (normalized amplitude of the scalar perturbation $A_s^2\equiv4/25\, P_s$). More precisely, we impose $P_s=2.45\times 10^{-9}$ at the pivot scale $k_0=0.002$ Mpc$^{-1}$ (Figs.~\ref{ratioconstraint} shows a constraint of the potential). Notice that at this large scale we expect the extreme slow-roll approximation to be valid. Indeed, these are the modes that exit the horizon at the very early time, deep enough in the inflationary era.
%%%%%%%%%%%%%%%%%%%%%%%%%%%%%%%%%%%%%%%%%%%%%
\begin{figure}[b]
\centering
  \includegraphics[width=0.35\textwidth]{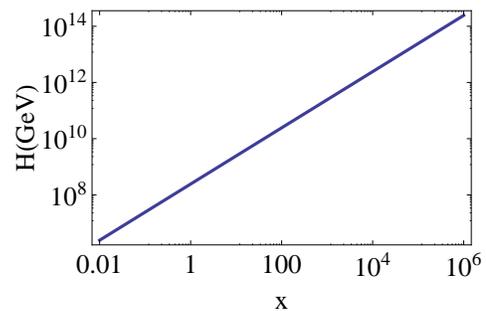}
	\caption{The Hubble parameter versus the dimensionless energy scale $x\equiv H/\mu$. Here we have fixed the parameter $\kappa_4\mu=10^{-10}$ as an example.}
	\label{H}
\end{figure}
%%%%%%%%%%%%%%%%%%%%%%%%%%%%%%%%%%%%%%%%%%%%%
%%%%%%%%%%%%%%%%%%%%%%%%%%%%%%%%%%%%%%%%%%%%%
\begin{figure*}
  \centering
  \subfloat[]{\label{Vgamma}\includegraphics[width=0.4\textwidth]{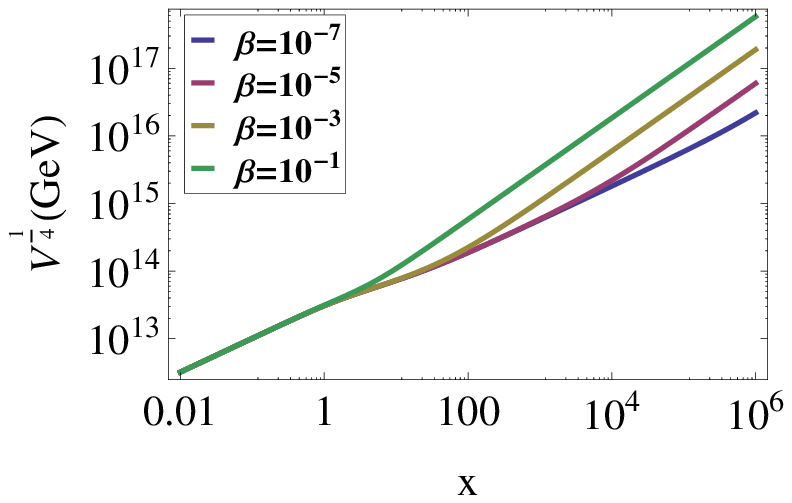}}
  \qquad
  \subfloat[]{\label{epsilongamma}\includegraphics[width=0.4\textwidth]{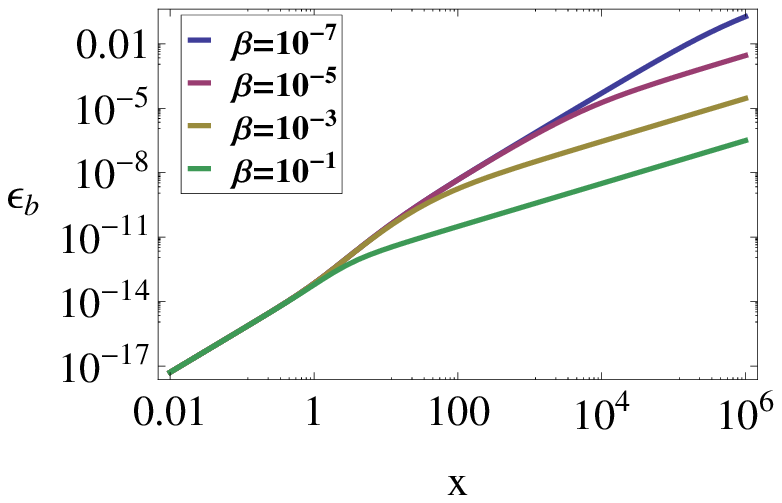}}
	\caption{The energy scale of the potential $V$ and the slow-roll parameter $\epsilon_b$ during extreme slow-roll inflation versus the dimensionless energy sale $x\equiv H/\mu$. Here we have fixed the parameter $\kappa_4\mu=10^{-10}$ as an example and the IG parameter, $\gamma=0.1$, and changed the GB parameter $\beta$ as shown on the plots. The brane tension can then be easily fixed through Eqs.(2.8) and (4.3).}
	\label{Vepsilongamma}
\end{figure*}
%%%%%%%%%%%%%%%%%%%%%%%%%%%%%%%%%%%%%%%
\begin{figure*}[!ht]
  \centering
  \subfloat[]{\label{Vbeta}\includegraphics[width=0.4\textwidth]{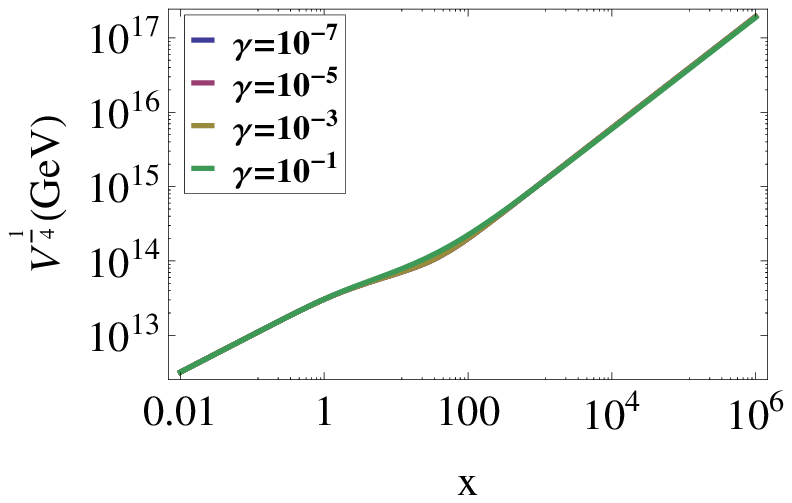}}
  \qquad
  \subfloat[]{\label{epsilonbeta}\includegraphics[width=0.4\textwidth]{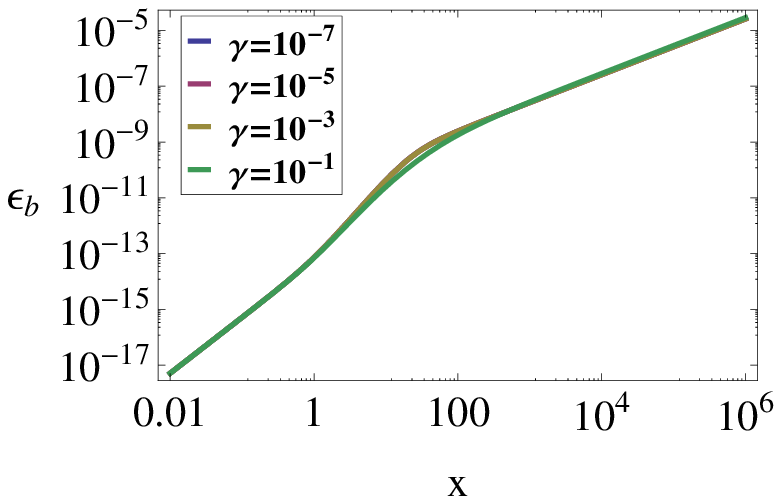}}
	\caption{The energy scale of the potential $V$ and the slow-roll parameter $\epsilon_b$ during extreme slow-roll inflation versus the dimensionless energy sale $x\equiv H/\mu$. Here we have fixed the parameter $\kappa_4\mu=10^{-10}$ as an example and the GB parameter, $\beta=10^{-3}$, and changed the IG parameter $\gamma$ as shown on the plots. The brane tension can then be easily fixed through Eqs. (2.8) and (4.3).}
	\label{Vepsilonbeta}
\end{figure*}
%%%%%%%%%%%%%%%%%%%%%%%%%%%%%%%%%%%%%%

Our constraints can be extended by considering the measure of the spectral index $n_s$ at the same pivot scale:
\begin{equation}
n_s-1\equiv\frac{d\ln A_s^2}{d\ln k},
\end{equation}
which can be calculated using the first two slow-roll parameters in the slow-roll approximation:
\begin{equation}
n_s\approx1-6\epsilon_b+2\eta_b,
\end{equation}
where the slow-roll parameters $\epsilon_b$ and $\eta_b$ are defined by
\begin{align}
\epsilon_b&\equiv-\frac{\dot{H}}{H^2}=\frac1{2\kappa_4^2}\left(\frac{V'}{V}\right)^2 C^1_{\alpha,\gamma}(x),\label{epsilon}  \\
\eta_b&\equiv \frac{V''}{3H^2}=\frac1{\kappa_4^2}\left(\frac{V''}{V}\right)C^2_{\alpha,\gamma}(x),\label{eta}
\end{align}
where $C^1_{\alpha,\gamma}(x)$ and $C^2_{\alpha,\gamma}(x)$ are the corrections to the standard 4d results given in the appendix \ref{appendix2} (see also Fig.~\ref{slowroll}). The corrections to the slow-roll parameters are given in Eqs.(\ref{C1})-(\ref{C2}). Those corrections depend on the GB and IG effects through the dimensionless parameters $\beta$ and $\gamma$. In addition, as can be seen from Eqs.(\ref{C1})-(\ref{C2}), the corrections $C^1_{\alpha,\gamma}$ and $C^2_{\alpha,\gamma}$ depend also on the ratio between the potential and the brane tension $V/\lambda$, on the one hand, and on the dimensionless rescaled and shifted Hubble parameter $\bar X$ of the normal branch, on the other hand. Nevertheless, $V/\lambda$ and $\bar X$ can be written as shown in Eq.(\ref{Xbar}) and Eq.(\ref{ratio}). Therefore, in summery the corrections to the slow-roll parameters depend exclusively on $\beta$, $\gamma$, and $x$. Notice that all the equations given in the appendix \ref{appendix2} already include the RS2 kind of fine-tuning condition (\ref{fine-tuning}) and the relation in Eq.(\ref{graviconst}).

We see that, from Fig.~\ref{slowroll}, the slow-roll parameters are strongly enhanced in the very high energy limit relative to the standard 4d results. Therefore, the slow-roll conditions $\epsilon_b\ll1, \,\,\eta_b\ll1$ cease to be valid at the very high energy limit for a given potential due to the GB effect. Since WMAP7 data suggest a nearly scale-invariant power spectrum at pivot scale \cite{Komatsu:2010fb} and we expect the extreme slow-roll approximation to be fulfilled at this large scale, the slow-roll conditions should be satisfied as well.

Furthermore, we can deduce the energy scale where inflation take place, i.e., we can obtain the range of the Hubble parameter and the inflaton potential. To do this we use the equation (\ref{ratio}) and for simplicity we fix the value of $\mu$ as shown in Figs.~\ref{H}, \ref{Vepsilongamma}, and \ref{Vepsilonbeta}. Following this procedure we can as well check the validity of the slow-roll conditions, which is indeed the case as shown on the right hand side plots of Figs.~\ref{Vepsilongamma} and \ref{Vepsilonbeta}.

For the slow-roll parameter $\epsilon_b$ on these plots, since the power spectrum of the scalar perturbation (\ref{As}) is constrained by the WMAP7 data, i.e., $P_s=2.45\times 10^{-9}$ at the pivot scale $k_0=0.002$ Mpc$^{-1}$, we can rewrite the standard 4d slow-roll parameter $\epsilon_{4D}\equiv1/2\kappa_4^2\cdot(V'/V)^2$ as follows combing the amplitude of the scalar perturbation Eq.(\ref{As}) with the conditions Eq.(\ref{fine-tuning}) and Eq.(\ref{graviconst}):
\begin{equation}
\epsilon_{4D}=\frac{(\kappa_4\mu)^2}{12\pi^2P_s}(1-\gamma)\left(\frac{3-\beta}{1+\beta}\right)\left(\frac V{\lambda}\right)G^2_{\alpha,\gamma}(x),\label{epsilon4d}
\end{equation}
where the ratio $V/\lambda$ can also be expressed as a function of the dimensionless energy scale $x$ through the  Eq.(\ref{ratio}). Thus from Eq.(\ref{epsilon}) and Eq.(\ref{epsilon4d}), we can rewrite the slow-roll parameter $\epsilon_b$ in terms of the dimensionless energy scale $x$ with three unknown parameters $\beta$, $\gamma$, and $\mu$, i.e.,  $\epsilon_b=\epsilon_{4D}\cdot C^1_{\alpha,\gamma}(x)$.

From Figs.~\ref{Vepsilongamma} and \ref{Vepsilonbeta} we see that the slow-roll parameter $\epsilon_b$ is very tiny for different values of $\beta$, $\gamma$, and $x$ as shown on these plots, which are consistent with the extreme slow-roll approximation, even though the energy scale of the potential $V^{1/4}$ can range widely for several energy scales. In addition, both the potential of the inflaton $V$ and the slow-roll parameter $\epsilon_b$ are more sensitive to the GB effect $\beta$ than the IG strength $\gamma$.

\section{Summary}

Brane inflation provides an interesting scenario for the early universe allowing us to explore the cosmological properties of higher-dimensional models. At such high energy scales, the non-conventional brane-world effects become dominant. Therefore, by investigating the scalar perturbations during inflation, we can see the modification from the brane effects relative to the standard general relativistic result. Here, we focus on the RS2-type brane-world modified by a GB correction term in the bulk as well as the strength of the IG effect on the brane. In order to compare with the RS2 model, we choose the normal branch for the cosmic evolution to compute the amplitude of the scalar perturbations, which reduces to the RS2 model in the absence of GB and IG corrections, in the slow-roll limit.

In such brane-world inflationary  model and in the extreme slow-roll limit, we assume that there is no scalar zero-mode contribution from the bulk, and moreover the massive scalar modes are too heavy to be excited during inflation \cite{Langlois:2000ns,BouhmadiLopez:2004ax,Dufaux:2004qs}. We can therefore safely disregard any extra scalar degree of freedom from the bulk and we can simply take into account the standard GR result for the amplitude of single field scalar perturbations.  Along this line of thought, in this paper we have calculated the corrections to the standard GR results for the scalar perturbations in a brane-world model with the curvature effects mentioned earlier. The amplitude of the scalar perturbations in the RS2 model is monotonically increasing with respect to the energy scale. However, unlike RS2 case the effect from the GB correction in an IG brane-world model is to decrease the amplitude of the scalar perturbations, and a similar result is obtained for the IG effect in a GB brane-world. Furthermore, in the high energy limit the perturbation will be highly suppressed by the GB effect.

Finally, we constrain the model using WMAP7 data. More precisely, we use the value of the power spectrum of the scalar perturbations $P_s=2.45\times 10^{-9}$ at a given pivot scale $k_0=0.002$ Mpc$^{-1}$. Our results correspond to: (i) a constraint of the potential as shown in Fig.~\ref{ratioconstraint}. (ii) deduction of the scale of inflation in terms of the dimensionless energy scale as shown in Figs.~\ref{Vepsilongamma} and \ref{Vepsilonbeta}. (iii) we check that our results are in full agreement with the extreme slow-roll approximation we used (see Figs.~\ref{Vepsilongamma} and \ref{Vepsilonbeta}).

\acknowledgments

M.B.L. is supported by the Spanish Agency ``Consejo Superior de Investigaciones Cient\'{\i}ficas" through JAEDOC064.
She also wishes to acknowledge the hospitality of LeCosPA Center at the National Taiwan University during the completion of part of this work and the support of the Portuguese Agency ``Fund\c{c}\~{a}o para a Ci\^{e}ncia e Tecnologia" through PTDC/FIS/111032/2009.
P.C. and Y.W.L. are supported by Taiwan National Science Council under Project No. NSC 97-2112-M-002-026-MY3 and by Taiwan’s National Center for Theoretical Sciences (NCTS). P.C. is in addition supported by US Department of Energy under Contract No. DE-AC03-76SF00515.
This work has been supported by a Spanish-Taiwanese Interchange Program with reference 2011TW0010 (Spain) and NSC 101-2923-M-002-006-MY3 (Taiwan)

\appendix
\section{Solutions of the Friedmann equation}\label{appendix1}
The general solutions of the cubic equation $X^3+a_2X^2+a_1X+a_0=0$ are \cite{Abramowitz}:
\begin{align}
&X_1=S_1+S_2-\frac{a_2}3\label{1stsol},   \\
&X_2=-\frac12(S_1+S_2)-\frac{a_2}3+\frac{i\sqrt{3}}2(S_1-S_2)\label{2ndsol},   \\
&X_3=-\frac12(S_1+S_2)-\frac{a_2}3-\frac{i\sqrt{3}}2(S_1-S_2)\label{3rdsol},
\end{align}
where
\begin{align}
 S_1=\left[R+\left(Q^3+R^2\right)^{\frac12}\right]^{\frac13}, \\ S_2=\left[R-\left(Q^3+R^2\right)^{\frac12}\right]^{\frac13},
\end{align}
with the definitions
\begin{equation}
R=\frac16(a_1a_2-3a_0)-\frac1{27}a^3_2,
\end{equation}
and
\begin{equation}
Q=\frac13a_1-\frac19a^2_2.
\end{equation}
And the discriminant $N=Q^3+R^2$ determines the number of the real solutions \cite{Abramowitz}:
\begin{align}
\textrm{(i)}&N>0:\,\,  \mbox{one real root and a pair of complex conjugate},\notag\\
\textrm{(ii)}&N=0:\,\,  \mbox{all roots real and at least two are equal},\notag\\
\textrm{(iii)}&N<0:\,\,  \mbox{all roots real (irreducible case)}\notag.
\end{align}

\section{Slow-roll parameters}\label{appendix2}
The explicit form of the corrections $C^1_{\alpha,\beta}(x)$ and $C^2_{\alpha,\beta}(x)$ to the standard 4d slow-roll parameters are shown below:
\begin{widetext}
\begin{align}
C^1_{\alpha,\gamma}(x)=&\frac83\frac{(1-\gamma)^2}{\gamma}\left(\frac{3-\beta}{1+\beta}\right)^2\left(\frac V{\lambda}\right)^2\bar{X}\left[(3\bar{X}+1)^2+(3b-1)\right]^{-1}\left\{\left[\frac34\left(\frac{\gamma}{1-\gamma}
\right)\left(\frac{1+\beta}{\beta}\right)\bar{X}\right]^2-1\right\}^{-2},\label{C1} \\
C^2_{\alpha,\gamma}(x)=&\frac23(1-\gamma)\left(\frac{3-\beta}{1+\beta}\right)\left(\frac V{\lambda}\right)\left\{\left[\frac34\left(\frac{\gamma}{1-\gamma}\right)\left(\frac{1+\beta}
{\beta}\right)\bar{X}\right]^2-1\right\}^{-1},\label{C2}
\end{align}
\end{widetext}
where the dimensionless parameter $\bar{X}$ is related to the Hubble parameter of the normal branch as discussed in Sec.~\ref{branches}, and the dimensionless parameter $b$ and energy density $\bar{\rho}$ can be rewritten as
\begin{align}
b=&\frac83\left(\frac{1-\gamma}{\gamma}\right)^2\frac{\beta(1-\beta)}{(1+\beta)^2}, \\
\bar{\rho}=&\frac{16}9\left(\frac{1-\gamma}{\gamma}\right)^2\left(\frac{\beta}{1+\beta}\right)^2+\notag\\
&\frac{32}{27}\left(\frac{1-\gamma}{\gamma}\right)^3\frac{\beta^2(3-\beta)}{(1+\beta)^3}\left(1+\frac{V}{\lambda}\right).
\label{rhobar2}
\end{align}
Notice that we have substituted the RS2 kind of fine-tuning condition Eq.(\ref{fine-tuning}) and the relation between $\kappa_4$ and $\kappa_5$ Eq.(\ref{graviconst}) in above results Eqs.(\ref{C1})-(\ref{rhobar2}). Moreover, the corrections to the standard 4d results Eq.(\ref{C1}) and Eq.(\ref{C2}) reduce to the RS2 model in the absence of GB and IG effects \cite{hep-ph/9912464},i.e., when $\alpha\rightarrow0$ and $\gamma\rightarrow0$:
\begin{align}
C^1_{\alpha,\gamma}\rightarrow C^1_{RS}=&\frac{4\lambda(\lambda+V)}{(2\lambda+V)^2}, \\
C^2_{\alpha,\gamma}\rightarrow C^2_{RS}=&\frac{2\lambda}{2\lambda+V}.
\end{align}

In addition, the ratio $V/\lambda$ and the dimensionless parameter $\bar{X}$ can be further expressed in terms of the dimensionless energy scale $x$:
\begin{align}
\bar{X}=&\frac43\left(\frac{1-\gamma}{\gamma}\right)\left(\frac{\beta}{1+\beta}\right)\sqrt{1+x^2},\label{Xbar}\\
\frac V{\lambda}=&\sqrt{1+x^2}\left(1+\frac{2\beta}{3-\beta}x^2\right)+\notag\\
&\frac23x^2\left(\frac{\gamma}{1-\gamma}\right)\left(\frac{1+\beta}{3-\beta}\right)-1.\label{ratio}
\end{align}

In the very low energy limit, i.e., $H\ll\mu$ or $x\rightarrow0$, the corrections $C^1_{\alpha,\gamma}(x)\sim1$ and $C^2_{\alpha,\gamma}(x)\sim1$, and therefore the slow-roll parameters $\epsilon_b$ and $\eta_b$ recover the standard 4d results at very low energy limit. By contrast, in the very high energy regime, i.e., $\mu\ll H$ or $x\rightarrow\infty$, the corrections $C^1_{\alpha,\gamma}(x)$ and $C^2_{\alpha,\gamma}(x)$ have the following approximation:
\begin{align}
C^1_{\alpha,\gamma}(x)\sim&\frac89(1-\gamma)\left(\frac{\beta}{1+\beta}\right)x,\label{C1approximation} \\
C^2_{\alpha,\gamma}(x)\sim&\frac83(1-\gamma)\left(\frac{\beta}{1+\beta}\right)x.\label{C2approximation}
\end{align}
From Eq.(\ref{C1approximation}) and Eq.(\ref{C2approximation}) we see that the corrections are monotonically enhanced at very high energy limit due to the GB effect even when the effect is tiny.


\begin{thebibliography}{99}
\bibitem{footnote2}
Please note that we have left the induced gravity parameter $\gamma$ arbitrary and not set to one as was done in Ref.~\cite{Kofinas:2003rz}.
\bibitem{footnote2-a}
Notice that we are not considering the limiting case corresponding to the Chern-Simons gravity because on that case a homogenous and isotropic brane cannot be embedded in the bulk \cite{Charmousis:2002rc}.
\bibitem{footnote3}
The solutions of Eq.(\ref{sqrt}) for the ``$-$'' sign are the same as those for Eq.(\ref{cubic}) with the change $\bar{X}\rightarrow\bar{X}=-8\alpha/3\gamma r \sqrt{H^2+\mu^2}$.
\bibitem{clarification}
The inflaton potential should in principle depend on the inflaton field. However, we are calculating the scalar perturbations within the extreme slow-roll regime which means that the inflationary phase of the brane is de Sitter-like; i.e. not only the kinetic term of the scalar field can be negleted in comparison with the scalar field potential  but also the potential can be assumed to be constant and therefore independent of the inflaton field as given in Eq. (4.8). In order to tackle the next order on the slow-roll approximation the dependence of the potential on the scalar field has to be assumed. This is technically much more involved and out of scope of the present paper as it would imply an analysis of the scalar pertubations in the bulk which would be coupled to those on the brane. We hope to tackle this issue on the near future.
\end{thebibliography}
\end{document}